\documentclass[12pt]{article} 
\usepackage[colorlinks, citecolor={blue}]{hyperref}
\usepackage{url}
\usepackage{amsfonts,amscd,amssymb}
\usepackage{amsthm,amsmath,natbib}
\usepackage{algorithm,algorithmicx,algpseudocode}
\usepackage{bm}
\usepackage{bbm} 
\usepackage[table]{xcolor}
\usepackage{verbatim}
\usepackage{graphicx}
\usepackage{setspace}
\usepackage{natbib}
\usepackage[margin=1in]{geometry}
\usepackage{enumitem}
\usepackage{listings}
\usepackage[textsize=tiny]{todonotes}
\usepackage{tikz}
\usepackage{etoolbox}
\usepackage{appendix}
\usepackage[format=plain,
font=it]{caption}
\usepackage{subcaption}
\usepackage{wrapfig}
\usepackage{xr}
\usepackage{booktabs}
\usepackage{multirow}
\usepackage{authblk}

\usetikzlibrary{matrix}
\usetikzlibrary{backgrounds}
\usetikzlibrary{calc}
\usetikzlibrary{arrows,shapes}
\usetikzlibrary{decorations}
\usetikzlibrary{decorations.pathmorphing}
\usetikzlibrary{fit}
\usetikzlibrary{decorations.pathreplacing}

\newtoggle{quickdraw}
\toggletrue{quickdraw} 

\definecolor{lightgrey}{rgb}{0.9,0.9,0.9}
\definecolor{darkgreen}{rgb}{0,0.3,0}

\definecolorset{rgb}{}{}{darkred,0.8,0,0;darkgreen,0,0.5,0;darkblue,0,0,0.5}

\renewcommand{\vec}[1]{\mathbf{#1}}

\newcommand{\latentdata}{\vec{x}}

\definecolor{trevorblue}{rgb}{0.330, 0.484, 0.828}
\definecolor{trevoryellow}{rgb}{0.829, 0.680, 0.306}

\title{Scalable Bayesian inference for self-excitatory stochastic processes applied to big American gunfire data}
\date{}

\author[1]{Andrew J.~Holbrook}
\author[2]{Charles E.~Loeffler}
\author[3]{Seth R.~Flaxman}
\author[1,4,5]{Marc A.~Suchard}
\affil[1]{Department of Biostatistics, University of California, Los Angeles}
\affil[2]{Department of Criminology, University of Pennsylvania}
\affil[3]{Department of Mathematics, Imperial College London}
\affil[4]{Department of Biomathematics, University of California, Los Angeles}
\affil[5]{Department of Human Genetics, University of California, Los Angeles}

\begin{document}

\maketitle

\begin{abstract}
The Hawkes process and its extensions effectively model self-excitatory phenomena including earthquakes, viral pandemics, financial transactions, neural spike trains and the spread of memes through social networks.  The usefulness of these stochastic process models within a host of economic sectors and scientific disciplines is undercut by the processes' computational burden: complexity of likelihood evaluations grows quadratically in the number of observations for both the temporal and spatiotemporal Hawkes processes.  We show that, with care, one may parallelize these calculations using both central and graphics processing unit implementations to achieve over 100-fold speedups over single-core processing. Using a simple adaptive Metropolis-Hastings scheme, we apply our high-performance computing framework to a Bayesian analysis of big gunshot data generated in Washington D.C.~between the years of 2006 and 2019, thereby extending a past analysis of the same data from under 10,000 to over 85,000 observations.  To encourage wide-spread use, we provide \textsc{hpHawkes}, an open-source \textsc{R} package, and discuss high-level implementation and program design for leveraging aspects of computational hardware that become necessary in a big data setting.
\end{abstract}

\clearpage

\section{Introduction}

The gun violence epidemic in the United States is associated with over 30,000 deaths each year and over 650,000 deaths in the past twenty \citep{causeofdeath2020}.  Although a serious problem, mass shootings only account for a small fraction of these deaths, while gun related homicides are most common in poor metropolitan areas \citep{bjerregaard1995gun,national2013priorities}. In 2005, for example, the highest per capita gun homicide rate in the country was 35.4 per 100,000 inhabitants in Washington D.C. \citep{crimeUS2005}.  Despite its massive scale, the nature of gun related violence and its impact on U.S.~public health remains poorly understood due, in part, to a paucity in the number of researchers focused on the field.  In 2013, there were only 20 academic researchers in the U.S. focusing on gun violence ``and most of them [were] economists, criminologists or sociologists'' \citep{wadman2013firearms}.  This dearth in public health experts studying gun violence is largely due to the 1996 Dickey Amendment prohibiting the Centers for Disease Control and Prevention (CDC) from promoting gun control \citep{wadman2013firearms,rubin2016tale}. Similarly, for researchers interested in studying gun violence, data availability has been a persistent issue \citep{national2005firearms,national2013priorities}. While jurisdictions routinely report incidents where individuals are killed using firearms, nonfatal and near miss incidents, which vastly outweigh fatal firearm incidents, have been much less reliably reported.

But there are two reasons for (tempered) hope that we will better understand gun violence in the future.  First, the federal budget for the 2020 fiscal year includes expenditures up to \$25 million dollars to be split between the CDC and the National Institutes of Health for research in the reduction of gun-related deaths and injuries, marking the first such expenditure since 1996 \citep{grisales_2019}. Second, new kinds of data that may shed light on the nature of gun violence have become publicly available within the past decade. Examples of this recent expansion in gun violence data availability include local police department open data portals, crowdsourced gun violence reporting systems and journalistic data initiatives.
 One new source of data, acoustic gunshot locator systems (AGLS; \cite{showen1997operational}) use a spatially distributed network of acoustic sensors to triangulate the locations of gunshots in space and time, thus overcoming the fact that the majority of gunshots go unreported to law enforcement \citep{mares2012evaluating}.   And so, a new challenge arises:  combining the massive scale of American gun violence with the fidelity of AGLS results in a potential deluge of big American gunfire data, and we must develop the computational and statistical techniques to effectively analyze them.

 Analysis of gun violence in the United States has long relied on a range of modeling approaches drawn from spatiotemporal statistics including classical Knox tests, K-functions, and more recent developments such as Gaussian processes \citep{ratcliffe2008near,flaxman2015machine}. Using these tests, scholars have sought to detect and study the degree of stability of gun violence clusters, sometimes referred to as gun violence hotspots. They have also explored whether gun violence diffuses in space and time as well as within social networks of susceptible places and populations and whether public health and law enforcement interventions designed to reduce the toll of gun violence are effective in diminishing its incidence. Examples of such interventions include violence interruption programs, focused deterrence initiatives as well as more traditional policing interventions. All of these rely on spatiotemporal measures to generate evidence of both theoretical and policy significance. Using these reliable methods, as they have been successfully used in other public health domains, scholars have learned that only some types of gun violence reliably cluster and that at least some violence can be disrupted \citep{park2019investigating}.

 At the same time, these implementations draw heavily on the availability of relatively sparse point process data or sensibly aggregated point process data to enable both inferential and predictive work. However, with the arrival of newer and higher resolution data sources such as AGLS, many oft-posed research questions need to be revisited in order to test whether the assumptions built into classical analyses hold up. Furthermore, some research questions that have been left unanswered due to the challenge of answering them using data measured with relatively low spatial and temporal resolution---to say nothing of missing data due to non-reporting---can now be explored. Key unresolved questions include the exact scales at which violence diffuses.

As a case study in the statistical analysis of big gunfire data, we consider the Washington D.C. ShotSpotter AGLS dataset \citep{petho_2013} consisting of over 85,000 potential gunfire events from 2006 to 2019.  A previous analysis of these data \citep{loeffler2018gun} restricts itself to a relatively small subset of around 9,000 events occurring in the years 2010 through 2012.  That same paper seeks to determine whether evidence exists for gun violence being contagious in the sense of bursts of diffusions through the urban landscape.   We follow \cite{loeffler2018gun} and model this contagiousness using the \emph{self-excitatory}  spatiotemporal Hawkes process \citep{reinhart2018review}, the computational complexity of which, unfortunately, scales quadratically in the number of observed events.  As a result, scaling model calculations to all 85,000 events is difficult, but we overcome this challenge with the aid of massive parallelization and cutting-edge computational hardware.

The temporal Hawkes process \citep{hawkes1971spectra,hawkes1971point,hawkes1972spectra} and its extensions are stochastic point processes that effectively model phenomena that are \emph{self-excitatory} in nature. Given an earthquake, we expect to observe aftershocks soon after and close to the epicenter; and a meme that is `going viral' triggers a cascade of `likes' that traverses the edges connecting a social network. Similarly, a diffusion of biological viruses across a human landscape also exhibits self-excitatory behavior, where an infected student or coworker often results in infected students or coworkers. Hawkes processes and extensions have successfully modeled earthquakes \citep{hawkes1973cluster,ogata1988statistical,zhuang2004analyzing}, viral memes \citep{yang2013mixture,mei2017neural}, neural activity \citep{linderman2014discovering,truccolo2016point,linderman2017bayesian}, viral epidemics \citep{kim2011spatio,meyer2014power,choi2015constructing,rizoiu2018sir,kelly2019real} and financial transactions \citep{embrechts2011multivariate,chavez2012high,hardiman2013critical,hawkes2018hawkes}.

Due to the wide, multi-sector use of the entire family of extended Hawkes process models, we believe that a demonstration of their natural parallelizability will be beneficial to theoreticians and practitioners alike.  Specifically, we use Bayesian inference \citep{rasmussen2013bayesian,linderman2014discovering} to learn posterior distributions of our spatiotemporal Hawkes process model parameters conditioned on tens of thousands of observed events.  Our simple Markov chain Monte Carlo (MCMC; \citet{metropolis1953equation,hastings1970monte}) algorithm requires repeated likelihood evaluations, each of which scales quadratically in computational complexity.  Overcoming this bottleneck in a big data setting is the chief contribution of our work.

A robust literature exists for parallel implementations in statistical computing: \cite{suchard2009many}, \cite{suchard2010understanding} and \cite{suchard2010some} perform optimization and Bayesian inference using graphics processing units (GPUs); \cite{lee2010utility} and \cite{zhou2010graphics} use the same hardware for sequential Monte Carlo and statistical optimization, respectively; and \cite{beam2016fast} apply GPUs to the evaluation of  the multinomial likelihood and its gradient.   More recently, \cite{warne2019acceleration} explore the use of central processing unit (CPU) based  single instruction-multiple data (SIMD) vectorization in various tasks within Bayesian inference, and  \citet{holbrook2019massive} use GPUs, multi-core CPUs and SIMD vectorization to accelerate MCMC for Bayesian multidimensional scaling with millions of data points.  In a similar manner, we develop a high-performance computing framework for scalable MCMC for the spatiotemporal Hawkes process  using  many-core GPU, multi-core CPU and SIMD vectorization based implementations.  To increase the impact of our work, we provide this high-performance computing framework as \textsc{hpHawkes}, a rudimentary, open-source \textsc{R} package freely available at  \url{https://github.com/suchard-group/hawkes}.

\section{Methods}

\subsection{Model}\label{sec:model}

\newcommand{\x}{\mathbf{x}}
\newcommand{\dd}{\mbox{d}}

The spatiotemporal Hawkes process describes the joint distribution of random variables $(\x,t)\in \mathbb{R}^D\times \mathbb{R}^+$ in space and time as an inhomogeneous Poisson process \citep{daley2003introduction,daley2007introduction} with intensity function
\begin{align*}
\lambda(\x,t) = \mu(\x,t) + \sum_{t_n<t} g(\x - \x_n, t - t_n)
\end{align*}
conditioned on observations $(\x_1,t_1)$, ..., $(\x_N,t_N)$.  In this formulation, $\mu(\cdot,\cdot)$ is the background or endemic rate, and $g(\cdot,\cdot)$ is the triggering function describing the self-excitatory nature of the process. We follow \cite{mohler2014marked} and \cite{loeffler2018gun} in the use of a triggering function that is exponential in time and Gaussian in space when modeling crime data:
\begin{align*}
\lambda(\x,t) = \mu(\x,t) + \frac{\theta}{\omega h^d} \sum_{t_n<t} e^{- \omega\, (t-t_n) }  \phi\left(\frac{\x-\x_n}{h}\right) \, .
\end{align*}
Parameters $\omega$, $h$ and $\theta$ are strictly positive, and we call $1/\omega$ and $h$ the temporal and spatial bandwidths belonging to the conditional rate function's self-excitatory term.  We further opt for a flexible Gaussian kernel smoother to model the background rate
\begin{align*}
\mu(\x,t) = \frac{\mu_0}{\tau_x^d\,\tau_t} \sum_{n=1}^N \phi\left(\frac{\x-\x_n}{\tau_x}\right) \cdot \phi\left(\frac{t-t_n}{\tau_t}\right)
\end{align*}
with $\tau_x$ and $\tau_t$ the spatial and temporal bandwidths corresponding to the endemic background rate.  Taken together, $\mu_0$ and $\theta$ describe the extent to which the process is self-excitatory in nature.  Denoting $\Theta = (\mu_0,\tau_x,\tau_t,\theta,\omega,h)$, the likelihood \citep{daley2003introduction} for data  $(\x_1,t_1)$, ..., $(\x_N,t_N)$ is
\begin{align*}
\mathcal{L}(\Theta)
= \exp \left( - \int_{\mathbb{R}^D} \int_0^{t_N} \lambda(\x,t) \, \dd t\, \dd\x  \right)  \prod_{n=1}^N \lambda(\x_n,t_n) := e ^{ - \Lambda(t_N) } \cdot \prod_{n=1}^N \lambda_n  \, .
\end{align*}
Our intensity function separates in space and time, so the integral $\Lambda(t_N)$ factorizes. The spatial integral is unity, and \cite{laub2015hawkes} (Section 4.2) demonstrate the closed-form solution to the self-excitatory component's temporal integral with exponential triggering function:
\begin{align*}
\Lambda(t_N) &= \mu_0 \sum_{n=1}^N \left(\Phi\left(\frac{t_N-t_n}{\tau_t} \right) -\Phi\left(\frac{-t_n}{\tau_t} \right) \right) - \theta \sum_{n=1}^N \left( e^{-\omega\, (t_N-t_n)} -1 \right)  \\ \nonumber
&= \sum_{n=1}^N \left(  \mu_0\left(  \Phi\left(\frac{t_N-t_n}{\tau_t} \right) -\Phi\left(\frac{-t_n}{\tau_t} \right)\right)- \theta \left( e^{-\omega\, (t_N-t_n)} -1 \right) \right)  \\
&:= \sum_{n=1}^N \Lambda_n
\end{align*}
Thus we are able to calculate the log-likelihood
\begin{align}\label{eq:likelihood}
\ell(\Theta) &= - \Lambda(t_N) + \sum_{n=1}^N\log \lambda_n   \\ \nonumber
&= \sum_{n=1}^N\Bigg\{ \log \Bigg[  \sum_{n'=1}^N \Bigg( \frac{\mu_0}{\tau_x^d\,\tau_t} \phi\left(\frac{\x_n-\x_{n'}}{\tau_x}\right) \cdot \phi\left(\frac{t_n-t_{n'}}{\tau_t}\right) \\ \nonumber
& \quad +\frac{\theta\, \mathcal{I}_{[t_{n'}<t_n]}}{\omega h^d}e^{- \omega\, (t_n-t_{n'})}   \phi\left(\frac{\x_n-\x_{n'}}{h}\right) \Bigg)\Bigg]   - \Lambda_n \Bigg\} \\ \nonumber
&:= \sum_{n=1}^N \left[
\log \left(  \sum_{n'=1}^N \lambda_{nn'} \right)  - \Lambda_n \right] := \sum_{n=1}^N \ell_n  \, ,
\end{align}
which we use for Bayesian inference in the context of a simple MCMC algorithm (Section \ref{sec:inference}).  The likelihood's double summation over indices $n$ and $n'$ results in $\mathcal{O}(N^2)$ computational complexity. We overcome this computational burden by  developing parallel implementations of likelihood calculations on cutting-edge computational hardware (Section \ref{sec:parallelization}).  We also develop parallel implementations to compute the vector of probabilities $\pi_n$ that each individual event generates from self-excitation rather than from the background process:
\begin{align}\label{eq:prob_se}
\pi_n = \frac{\lambda_n - \mu_n}{\lambda_n} := \frac{\xi_n}{\lambda_{n}} \, ,
\end{align}
where $\xi_n$ denotes the self-excitatory component of rate $\lambda_n$. For each $n$, $\pi_n$ is a function of all $N-1$ other observations, so computing the entire vector is $\mathcal{O}(N^2)$.  Moreover, each $\pi_n$ is a function of $\Theta$, and we take the posterior distribution of each $\pi_n$ to be a key interpretable of our analysis. Given an MCMC sample $\Theta^{(1)}, \dots, \Theta^{(S)}$, obtaining a posterior sample $\pi^{(s)}_n$ for $n=1, \dots, N$ and $s=1,\dots,S$ is $\mathcal{O}(N^2S)$, again necessitating the cutting-edge computational hardware of Section \ref{sec:parallelization}.

To facilitate comparisons with \citet{loeffler2018gun}, we follow their specification and equip $\mu_0$ and $\theta$ with truncated normal priors with a lower bound of 0 and standard deviations of 1 and 10, respectively.  We lend truncated normal priors to $\omega$ and $1/h$ again with a lower bound of 0 and with a standard deviation of 10 for both.  Finally, we also follow that paper in setting the background rate's temporal and spatial lengthscales $\tau_t$ and $\tau_x$ to be 14 days and 1.6 kilometers.

 \subsection{Inference}\label{sec:inference}

Algorithm \ref{alg:mh} describes the simple, adaptive Metropolis-Hastings algorithm \citep{haario2001adaptive,roberts2009examples} with random scan univariate proposals we use to generate posterior realizations for $\omega$, $h$, $\theta$ and $\mu_0$.   Of the different algorithms described in the extensive adaptive MCMC literature, some of the simplest work by tuning the proposal distribution to obtain a target acceptance rate \cite{roberts2009examples}.  Following \citep{gelman1996efficient} we target an acceptance rate of 0.44 (Algorithm \ref{alg:mh}, Step 6d) for each of our four univariate proposals. We accomplish this while guaranteeing the \emph{diminishing adaptation} criterion of  \citet{roberts2007coupling} by increasing adaptation intervals at a super-linear rate (Algorithm \ref{alg:mh}, Step 6l).  For any interesting posterior distribution conditioned on even moderately sized data, the algorithm's computational bottleneck is the calculation of the likelihood function in Step 5a.  For most models belonging to the Hawkes process family, the computational complexity of this step is quadratic in the number of observations ($\mathcal{O}(N^2)$), and for our specific model this fact arises from the double summation of Equation \eqref{eq:likelihood}.  In the following section, we discuss the multiple parallelization strategies we use to overcome this rate-limiting step.

\newcommand{\bb}{\mathbf{b}}
\newcommand{\vv}{\mathbf{v}}
\newcommand{\aaa}{\mathbf{a}}
\newcommand{\adapt}{\mathbf{l}}

\newcounter{algsubstate}
\renewcommand{\thealgsubstate}{\alph{algsubstate}}
\newenvironment{algsubstates}
{\setcounter{algsubstate}{0}%
	\renewcommand{\State}{%
		\stepcounter{algsubstate}%
		\Statex {\footnotesize\thealgsubstate:}\space}}
{}

\begin{algorithm}[!ht]
	\caption{A simple adaptive Metropolis-Hastings algorithm: \\
\textit{Produces a Markov chain of $D$-dimensional states $\Theta^{(s)}>\mathbf{0}$  for $s=1,\dots,S$ with target density $p(\cdot)$.
	Following initialization, each iteration uses a random scan to generate a univariate proposal, accepts or rejects that proposal using a Metropolis-Hastings accept-reject step and updates the $D$ univariate proposal distributions with decreasing regularity.
	 We use 0.44 as target acceptance rate following \citet{gelman1996efficient}.}
} \label{alg:mh}
	\begin{algorithmic}[1]
    \State Initialize algorithmic quantities:
    \begin{algsubstates}
	\State  \hspace{1em}    Markov chain state counter $s \gets 1$
	\State  \hspace{1em} Markov chain state $\Theta^{(s)} = (\theta^{(s)}_1, \dots, \theta^{(s)}_D) \gets \Theta > \mathbf{0}$
	\State  \hspace{1em} adaptation interval bounds $\bb = (b_1,\dots,b_D) \gets (5, \dots, 5)$
	\State \hspace{1em}  adatation interval counters $\adapt = (l_1,\dots,l_D) \gets (0, \dots, 0)$
	\State  \hspace{1em}  acceptance counter $\aaa = (a_1,\dots,a_D) \gets (0, \dots, 0)$
	\State \hspace{1em} truncated normal proposal standard deviations $\vv = (v_1,\dots,v_D) \gets (1, \dots, 1)$
    \end{algsubstates}
	\For{s in $1:S$}
	\State Randomly select $d$th parameter to update: $d \sim$ Uniform$(1, \dots, D)$
	\State Generate proposal state $\Theta^* \sim q(\Theta^*|\Theta^{(s)})$:
	\begin{algsubstates}
		\State \hspace{1em} $\theta^{*}_d \sim$ Normal$(\theta^{(s)}_d,v_d)$ $\cdot$ I$(\theta^{*}_d>0)$
		\State \hspace{1em} $\Theta^* \gets (\theta^{(s)}_1,\dots, \theta^{*}_d, \dots,  \theta^{(s)}_D)$
	\end{algsubstates}
	\State Metropolis-Hastings accept-reject step:
	\begin{algsubstates}
		\State \hspace{1em} Calculate acceptance criterion: $r_1 \gets \left(p(\Theta^*) \, q(\Theta^{(s)}|\Theta^*) \right) /   \left(p(\Theta^{(s)}) \, q(\Theta^{*}|\Theta^{(s)}) \right) $
		\State \hspace{1em} Generate cut-off variable $u \sim$ Uniform$(0,1)$
		\State \hspace{1em} \textbf{if} $u < r_1 $ \textbf{then}
		\State \hspace{2em} $\Theta^{(s+1)} \gets \Theta^{*}$
		\State \hspace{2em} $a_d \gets a_d  + 1$
		\State \hspace{1em} \textbf{else}
		\State \hspace{2em}  $\Theta^{(s+1)} \gets \Theta^{(s)}$
		\State \hspace{1em} \textbf{end if-else}
	\end{algsubstates}
 	\State Update adaptation parameters:
	\begin{algsubstates}
	\State \hspace{1em} Increment adaptation interval counter $l_d \gets l_d + 1$
	\State \hspace{1em} \textbf{if} $l_d = b_d$  \textbf{then}
	\State \hspace{2em} Calculate proportion of acceptances $r_2 \gets a_d / b_d$
	\State \hspace{2em} Calculate adaptation ratio $r_3 \gets r_2  / 0.44$
	\State \hspace{2em} \textbf{if} $r_3>2$  \textbf{then}
	\State \hspace{3em} $r_3 \gets 2$
	\State \hspace{2em} \textbf{end if}
	\State \hspace{2em} \textbf{if} $r_3<0.5$  \textbf{then}
	\State \hspace{3em} $r_3 \gets 0.5$
	\State \hspace{2em} \textbf{end if}
	\State \hspace{2em} Update proposal standard deviation $v_d \gets r_3 \cdot v_d$
	\State \hspace{2em} Increase adaptation interval bound $b_d \gets b_d^{1.1}$
	\State \hspace{2em} Reset adaptation interval counter $l_d \gets 0$
	\State \hspace{2em} Reset acceptance counter $a_d \gets 0$
	\State \hspace{1em} \textbf{end if}
	\end{algsubstates}

	\EndFor
	\end{algorithmic}

\end{algorithm}

\newcommand{\transformR}{r}
\newcommand{\transformCDF}{c}
\newcommand{\threadsPerBlock}{B}

\algblock{ParFor}{EndParFor}
\algnewcommand\algorithmicparfor{\textbf{parfor}}
\algnewcommand\algorithmicpardo{\textbf{do}}
\algnewcommand\algorithmicendparfor{\textbf{end\ parfor}}
\algrenewtext{ParFor}[1]{\algorithmicparfor\ #1\ \algorithmicpardo}
\algrenewtext{EndParFor}{\algorithmicendparfor}

\begin{algorithm}[!ht]
	\caption{Parallel computation of Hawkes process likelihood: \\
		\emph{Uses multiple central processing unit (CPU) cores along with loop vectorization to compute log-likelihood.  For double-precision floating point, the algorithm uses either SSE or AVX vectorization to make $j=2$ or $4$ long jumps and cut loop iterations by one-half or three-fourths, respectively. Here, $B$ is the number of CPU threads available.   Symbols $\ell$, $\lambda$ and $\Lambda$ appear in Equation \eqref{eq:likelihood}.}
	}\label{alg:lik2}
	\begin{algorithmic}[1]

		\ParFor{ $b \in \{1,\dots,B\}$ }
		\State $\ell_b \gets 0$
		\If{$b\neq B$}
		\State $Upper \gets b \lfloor  N/B  \rfloor$
		\Else
		\State $Upper \gets  \lceil  N/B  \rceil$
		\EndIf
		\For{$n' \in  \{ (b-1)\lfloor N/B  \rfloor + 1,  \dots,Upper \}$}
		\State copy $\x_{n'}$, $t_{n'}$ to cache
		\State $\lambda_{n'} \gets \mathbf{0}$ \Comment{vector of length j}
		\State $n \gets 1$
		\While{ $n < N$ }
		\State $j \gets \min(j,N-n)$
		\State copy $\x_{n:(n+j)}$, $t_{n:n+j}$ to cache
		\State $\Delta_{n'n}:\Delta_{n'n:(n+j-1)} \gets (\latentdata_{n'} - \latentdata_{n}):(\latentdata_{n'} - \latentdata_{n+j-1})$ \Comment{vectorized subtraction}
		\State calculate $\delta_{n'n}:\delta_{n'(n+j-1)}$  \Comment{vectorized multiplication, see Algorithm \ref{alg:lik}}
		\State calculate $\lambda_{n'n}:\lambda_{n'(n+j-1)}$  \Comment{vectorized evaluation, see Algorithm \ref{alg:lik}}
		\State $\lambda_{n'} \gets \lambda_{n'} +  \lambda_{n'n}:\lambda_{n'(n+j-1)}$		\Comment{vectorized addition}
		\State $n \gets n + j$
		\EndWhile
		\State $\ell_b \gets \ell_b +  \log\left( \sum \{ \lambda_{n'}\}\right) - \Lambda_{n'}$
		\EndFor
		\EndParFor
		\State $\ell(\Theta) \gets \sum_{b} \ell_b$

	\end{algorithmic}
\end{algorithm}

\begin{algorithm}[!ht]
	\caption{Parallel computation of Hawkes process likelihood: \\
		\emph{Calculates the log-likelihood with multiple levels of parallelization on gpraphics processing unit (GPU).   In practice, we specify $B=128$ to be the the size of the GPU work groups.  Symbols $\ell$, $\lambda$ and $\Lambda$ appear in Equation \eqref{eq:likelihood}.}
	}\label{alg:lik}
	\begin{algorithmic}[1]
		\State Calculate observation-specific contributions to likelihood $\ell_n$:
		\begin{algsubstates}
			\State	\hspace{1em}	\textbf{parfor} $n \in \{1,\dots,N\}$ \textbf{do}
			\State\hspace{2em} copy $\x_n$, $t_n$ to local \Comment{$B$ threads}
			\State\hspace{2em}	\textbf{parfor} $N'\in \{1,\dots,\lfloor N/B\rfloor\}$ \textbf{do}
			\State\hspace{3em} $n' \gets N'$
			\State\hspace{3em} $\lambda_{nN'} \gets 0$
			\State\hspace{3em}\textbf{while} $n' < N$ \textbf{do}
			\State\hspace{4em} copy $\x_{n'}$, $t_{n'}$ to local \Comment{$B$ threads}
			\State\hspace{4em} $\Delta_{nn'} \gets \latentdata_n - \latentdata_{n'}$ \Comment{vectorized subtraction}
			\State\hspace{4em} calculate $\delta_{nn'} = \sqrt{\sum \Delta_{nn'}\circ \Delta_{nn'}}$  \Comment{vectorized multiplication}
			\State\hspace{4em} $\lambda_{nN'} \gets \lambda_{nN'}  +\lambda_{nn'}$ \Comment{$\lambda_{nn'}$ a function of $\delta_{nn'}$, $t_n$ and $t_{n'}$}
			\State\hspace{4em} $n' \gets n' + B$
			\State\hspace{3em}	\textbf{end while}
			\State\hspace{2em}\textbf{end parfor}
			\State\hspace{2em} $\lambda_n\gets \sum_{N'} \lambda_{nN'} $   \Comment{binary tree reduction on chip}
			\State\hspace{2em} $\ell_n \gets \log \lambda_n - \Lambda_n$
			\State\hspace{1em}	\textbf{end parfor}
		\end{algsubstates}

		\State Sum up all $N$ observation-specific contributions $\ell_n$:
		\begin{algsubstates}
			\State \hspace{1em}\textbf{parfor} $N'\in \{1,\dots,\lfloor N/B\rfloor\}$ \textbf{do}
			\State \hspace{2em}$n' \gets N'$
			\State \hspace{2em}$\ell_{N'} \gets 0$
			\State	\hspace{2em}\textbf{while} $n' < N$  \textbf{do}
			\State \hspace{3em}copy $\ell_{n'}$ to local \Comment{$B$ threads}
			\State \hspace{3em}$\ell_{N'} \gets \ell_{N'}  + \ell_{n'}$
			\State \hspace{3em}$n' \gets n' + B$
			\State\hspace{2em}\textbf{end while}
			\State \hspace{1em}\textbf{end parfor}
			\State \hspace{1em}$\ell(\Theta) \gets \sum_{N'} \ell_{N'} $   \Comment{binary tree reduction on chip}
		\end{algsubstates}

	\end{algorithmic}
\end{algorithm}

\begin{algorithm}[!ht]
	\caption{Parallel computation of self-excitatory probabilities: \\
		\emph{Uses multiple central processing unit (CPU) cores along with loop vectorization to compute $N$ self-excitatory probabilities $\pi_n$.  For double-precision floating point, the algorithm uses either SSE or AVX vectorization to make $j=2$ or $4$ long jumps and cut loop iterations by one-half or three-fourths, respectively. Here, $B$ is the number of CPU threads available.   Symbols $\pi_n$, $\mu_n$ and $\xi_n$ appear in Equation \eqref{eq:prob_se}.}
	}\label{alg:pi2}
	\begin{algorithmic}[1]

		\ParFor{ $b \in \{1,\dots,B\}$ }
		\If{$b\neq B$}
		\State $Upper \gets b \lfloor  N/B  \rfloor$
		\Else
		\State $Upper \gets  \lceil  N/B  \rceil$
		\EndIf
		\For{$n' \in  \{ (b-1)\lfloor N/B  \rfloor + 1,  \dots,Upper \}$}
		\State copy $\x_{n'}$, $t_{n'}$ to cache
		\State $\mu_{n'} \gets \mathbf{0}$ \Comment{vector of length j}
		\State $\xi_{n'} \gets \mathbf{0}$ \Comment{vector of length j}
		\State $n \gets 1$
		\While{ $n < N$ }
		\State $j \gets \min(j,N-n)$
		\State copy $\x_{n:(n+j)}$, $t_{n:n+j}$ to cache
		\State $\Delta_{n'n}:\Delta_{n'n:(n+j-1)} \gets (\latentdata_{n'} - \latentdata_{n}):(\latentdata_{n'} - \latentdata_{n+j-1})$ \Comment{vectorized subtraction}
		\State calculate $\delta_{n'n}:\delta_{n'(n+j-1)}$  \Comment{vectorized multiplication, see Algorithm \ref{alg:lik}}
		\State calculate $\mu_{n'n}:\mu_{n'(n+j-1)}$
		\State calculate $\xi_{n'n}:\xi_{n'(n+j-1)}$
		\State $\mu_{n'} \gets \mu_{n'} +  \mu_{n'n}:\mu_{n'(n+j-1)}$		\Comment{vectorized addition}
		\State $\xi_{n'} \gets \xi_{n'} +  \xi_{n'n}:\xi_{n'(n+j-1)}$		\Comment{vectorized addition}
		\State $n \gets n + j$
		\EndWhile
		\State $\pi_{n'} \gets \xi_{n'} / (\mu_{n'} +\xi_{n'})$              \Comment{vectorized addition, division and assignment}
		\EndFor
		\EndParFor
	\end{algorithmic}
\end{algorithm}

\begin{algorithm}[!ht]
	\caption{Parallel computation of self-excitatory probabilities: \\
		\emph{Calculates $N$ self-excitatory probabilities $\pi_n$ from single parameter value $\Theta$ with multiple levels of parallelization on gpraphics processing unit (GPU).   In practice, we specify $B=128$ to be the the size of the GPU work groups.  Symbols $\pi_n$, $\mu_n$ and $\xi_n$ appear in Equation \eqref{eq:prob_se}.}
	}\label{alg:pi}
	\begin{algorithmic}[1]
			\ParFor{$n \in \{1,\dots,N\}$ }
			\State copy $\x_n$, $t_n$ to local \Comment{$B$ threads}
			\ParFor{ $N'\in \{1,\dots,\lfloor N/B\rfloor\}$ }
			\State $n' \gets N'$
			\State $\mu_{nN'} \gets 0$
			\State $\xi_{nN'} \gets 0$
			\While{ $n' < N$ }
			\State copy $\x_{n'}$, $t_{n'}$ to local \Comment{$B$ threads}
			\State $\Delta_{nn'} \gets \latentdata_n - \latentdata_{n'}$ \Comment{vectorized subtraction}
			\State calculate $\delta_{nn'} = \sqrt{\sum \Delta_{nn'}\circ \Delta_{nn'}}$  \Comment{vectorized multiplication}
			\State $\mu_{nN'} \gets \mu_{nN'}  +\mu_{nn'}$ \Comment{$\mu_{nn'}$ a function of $\delta_{nn'}$, $t_n$ and $t_{n'}$}
			\State $\xi_{nN'} \gets \xi_{nN'}  +\xi_{nn'}$ \Comment{$\xi_{nn'}$ a function of $\delta_{nn'}$, $t_n$ and $t_{n'}$}
			\State $n' \gets n' + B$
            \EndWhile
			\EndParFor
			\State $\mu_n\gets \sum_{N'} \mu_{nN'} $   \Comment{binary tree reduction on chip}
			\State $\xi_n\gets \sum_{N'} \xi_{nN'} $   \Comment{binary tree reduction on chip}
			\State $\pi_n \gets \xi_n / (\mu_n +\xi_n)$
			\EndParFor
	\end{algorithmic}
\end{algorithm}

\subsection{Parallelization}
\label{sec:parallelization}

\newcommand{\blockSize}{B}

To parallelize the Hawkes process likelihood of Equation \eqref{eq:likelihood} and circumvent its $\mathcal{O}(N^2)$ computational complexity, we take a hardware oriented approach that uses four broad rules-of-thumb \citep{holbrook2019massive}.  \emph{First} and most importantly, we design our code to assign calculations of stereotyped and ostensibly independent terms to independent cores.  As such, we target the $N^2$ $\lambda_{nn'}$ terms of Equation \eqref{eq:likelihood} for simultaneous processing insofar as the hardware supports. \emph{Second}, we identify rate-limiting floating point calculations and perform them in parallel across vectors of inputs, thus providing an additional level of parallelization over and beyond the use of multiple cores.  For our model, the rate-limiting floating point calculations occur in the evaluation of $\exp(\cdot)$ in the individual $\lambda_{nn'}$s.  \emph{Third}, when calculations require the use of individual data multiple times, we store these data so as to encourage fast reuse.   For example, the calculation of $\lambda_{n}$ requires the evaluation of $N$ $\lambda_{nn'}$ terms, each of which depends on $\x_n$ and $t_n$.  \emph{Fourth}, we avoid costly storage of intermediate terms such as the individual $\lambda_{nn'}$ within our calculations and only store their running sum.

Different kinds of computational hardware capitalize on and facilitate these general strategies to different degrees.  Cluster computing scales to 1000s of CPUs connected by Ethernet or Infiniband networks, each CPU having its own random access memory (RAM).    The scale of such a cluster is undercut, however, by latency arising from communication between cluster nodes.  If one divides a computing task into two parts, the first being parallelizable and having sequential cost $c_0$, the second being non-parallelizable and having cost $c_1$, then one can accelerate compute time by sharing $c_0$ between $\nu$ nodes.  Unfortunately, Amdahl's law \citep{amdahl1967validity} says that the resulting wall time $c$ exhibits the bound
\begin{align*}
c \geq c_0/\nu + c_1
\end{align*}
on account of latency arising from parallel tasks finishing at different times and additional communication between nodes.  Indeed, for iterative algorithms such as MCMC, the lower-bound on $c$ becomes worse for every increasing iteration.  Such inefficiencies often result in diminishing returns for large clusters, which can require significant financial investments nonetheless \citep{suchard2010understanding}.

Given the latencies arising from iterative algorithms on large distributed-computing environments, we focus on the use of less expensive and more widely owned computing hardware to parallelize the evaluation of $\ell(\Theta)$, the bottleneck of our MCMC algorithm.  First, we use the multiple cores and SIMD vectorization supported by most modern CPUs that are available in standard desktop computers.  Second, we use the thousands of cores available in contemporary general purpose GPUs to achieve massive parallelization.  Specifically, we must use this hardware to parallelize the many \emph{transformations} and \emph{reductions} implied by Equation \eqref{eq:likelihood}.  For a fixed index $n$, reading $\x_n$, $t_n$, $\x_{n'}$ and $t_{n'}$ from global memory and evaluating $\lambda_{nn'}$ is a transformation.   Thus we require $N$ transformations to compute the $N$ terms within the inner summation of Equation \eqref{eq:likelihood}.  Following these transformations, a reduction maps from the individual $\lambda_{nn'}$s to their sum $\lambda_{n}$.  A further transformation reads $t_N$, $\x_n$ and $t_n$ from memory, computes $\Lambda_n$ from them and adds $\log(\lambda_n)$ and $\Lambda_n$ to obtain $\ell_n$.  A final reduction sums over all $N$ $\ell_n$ to obtain the likelihood $\ell(\Theta)$.  Regardless of the hardware type, we attack these transformation-reductions with the same general principles: we perform rate-limiting floating point operations such as those involved in the evaluation of $\lambda_{nn'}$ in parallel; we keep data in fast access memory when we require reuse (notice how $\x_n$ and $t_n$ appear in both transformations); and we use running summations to  avoid costly reading and writing of intermediate values such as $\lambda_{nn'}$.

\subsubsection{Multi-core CPUs}

Contemporary desktops and servers have sockets for as many as 8 CPU chips. These CPU chips contain 1 to 72 independent processing units called cores, each of which can perform different operations in parallel, and each chip contains three (or more) levels  of memory cache, L1, L2 and L3, that balance the rate of data transfer or \emph{memory bandwidth} with the amount of data storage available.  Typically, each core has its own L1 and L2 cache, where L1 has higher memory bandwidth but less storage than L2.   Cores on the same chip usually share L3 cache, which has even less memory bandwidth and even more storage than L2.  A memory bus connects on-chip cache to RAM, the bandwidth of which is significantly smaller than the total rate of numerical operations across cores.  In a big data setting, memory bandwidth becomes a bottleneck for even the most numerically intensive tasks.

Many programming languages contain software libraries that enable the computational statistician to communicate with a computer's operating system and coordinate the behavior of multiple cores in the performance of independent tasks.  We use \textsc{Threading Building Blocks} (\textsc{TBB}) \citep{Reinders:2007:ITB:1461409}, an open-source and cross-platform  \textsc{C++} library, for multi-core parallelization, and the \textsc{R} package \textsc{RcppParallel} makes \textsc{TBB} available to  \textsc{R} developers \citep{RcppParallel}.  These packages help to parallelize the transformation-reductions of Equation \eqref{eq:likelihood} by partitioning the task into $T$ threads, for $T$ less than or equal to the total number of cores of the multi-core environment.  Each thread is limited in the rate at which it performs the rate-limiting floating-point operations but has fast  and unimpeded access to L1 and L2 caches.  Specifically, we use \textsc{TBB} to assign calculation of elements $\lambda_{nn'}$, $n'=1,\dots,N$ to the same thread, so that a thread loads $\x_n$ and $t_n$ to an on-chip register for reuse $N$ times.   The same thread obtains $\lambda_n$ with a running summation of the $\lambda_{nn'}$ that avoids storage of intermediate values.  After computing $\lambda_n$, the exact same thread computes a partial sum from a subset of the $\ell_n$s and writes the partial sum to RAM.  Finally, a single thread sums the $T$ partial sums in a fast serial reduction. Algorithm \ref{alg:lik2} combines this multi-core implementation with within-core vectorization.  Algorithm \ref{alg:pi2} is similar to Algorithm \ref{alg:lik2} and computes the vector of self-excitatory probabilities $\pi_n$.

\subsubsection{Within-core vectorization}

One can further accelerate multi-core CPU processing with the aid of vector or SIMD processing \citep{warne2019acceleration,holbrook2019massive}, in which the CPU simultaneously applies a single set of instructions to data stored consecutively in an extended-length register.  For Intel x86 hardware, streaming SIMD extensions (SSE), advance vector extensions (AVX) and AVX-512 support vector operations on 128, 256 and 512 bit extended registers, respectively. For floating point operations in 64 bit double precision, this amounts to 2-fold, 4-fold and 8-fold theoretical speedups for SSE, AVX and AVX-512, although such performance gains rarely manifest in practice.  Whereas many computational statisticians know about multi-core processing, there is little mention of SIMD parallelization in the literature.  That said, some \textsc{R} wrapper packages such as \textsc{RcppXsimd} and \textsc{RcppNT2} \citep{RcppNT2} are becoming available and making it possible for \textsc{R} developers to employ SIMD intrinsics.

 We leverage SIMD parallelization by vectorizing or \emph{unrolling} loops within each thread and applying the entire loop body to an entire SIMD extended register at each iteration.  For AVX computing in double precision, each iteration of the unrolled loop corresponds to 4 iterations of the original loop.  This strategy benefits from efficient reading from and writing to consecutive memory locations and simultaneous evaluation of rate-limiting floating point operations.  The use of an instruction-level program profiler reveals that the rate-limiting step in our likelihood calculations is the evaluation of $\exp(\cdot)$ within the inner summation of Equation \eqref{eq:likelihood}.  Using AVX, for example, one evaluates $\exp(\cdot)$ over four doubles simultaneously and achieves a greater than 2-fold speedup.  With less impact on compute performance, we also vectorize the distance calculations between all pairs of location vectors $\x_n$ and $\x_{n'}$ \citep{holbrook2019massive}.

\subsubsection{Many-core GPUs}

GPUs contain hundreds to thousands of cores, but, unlike the independent cores of a CPU, small work-groups of GPU cores must execute the same instruction sets simultaneously though on different data.  In this respect, GPU based parallelization may be thought of as SIMD on a massive scale, leading Nvidia to coin the term SIMT (single instruction, multiple threads) \citep{lindholm2008nvidia}.  In this setup, communication between threads within the same work-group happens extremely quickly via shared on-chip memory, and scheduling a massive number of threads actually hides latencies arising from off-chip memory transactions because of the dynamic and simultaneous loading and off-loading of the many tasks.  In part, this is because of the GPU's massively parallel architecture. In part, this is because contemporary general purpose GPUs have small memory cache but high memory bandwidth, making them ideal tools for performing a massive number of short-lived, cooperative threads.

The likelihood evaluation first involves $N$ independent transformation-reductions, one to obtain each $\lambda_{n}$.  We generate $T=N\times B$ threads on the GPU and use work groups of $B$ threads to compute each of the $N$ $\lambda_{n}$.  Each thread uses a while-loop across indices $n'$ to compute $\lceil N/B \rceil$ $\lambda_{nn'}$s and keeps a running partial sum.  After the threads obtain $B$ partial sums, they work together in a final binary reduction to obtain $\lambda_n$.  The binary reduction is fast, with $\mathcal{O}(\log B)$ complexity and represents an additional speedup beyond massive parallelization.  After computing all $N$ $\lambda_n$s, a summation proceeds in the exact same manner.  The GPU uses massive parallelization to avoid the cost of the rate-limiting floating-point computation in $\exp(\cdot)$.   High memory-bandwidth allows for fast transfer to and from each working group, and, in turn, each work group shares its own fast access memory that facilitates rapid communication between member threads.  We use the Open Computing Language (\textsc{OpenCL}) to write our GPU code.  In \textsc{OpenCL}, write functions called kernels, and the library assigns them to each work group separately for parallel execution.  To evaluate the likelihood, we write one kernel for the work groups that compute the $\ell_n$s and one kernel for those that sum the $N$ $\ell_n$s.  These details culminate in Algorithm \ref{alg:lik}. Algorithm \ref{alg:pi} is similar to Algorithm \ref{alg:lik} and computes the vector of self-excitatory probabilities $\pi_n$.

\subsection{Software availability}\label{sec:software}

In writing this paper, we have developed \textsc{hpHawkes} \url{https://github.com/suchard-group/hawkes}, an open-source \textsc{R} package that enables massively parallel implementations of spatiotemporal Hawkes processes in a big data setting.   This package relies on \textsc{Rcpp} \citep{eddelbuettel2011rcpp} to build and interface with a \textsc{C++} library that uses \textsc{OpenCL} and \textsc{TBB} frameworks for parallelization on GPUs and CPUs, respectively. To enable within-core vectorization, \textsc{hpHawkes} accesses SIMD intrinsics via \textsc{RcppXsimd} \citep{holbrook2019massive}, an \textsc{R} package that itself uses \textsc{Rcpp} to access the \textsc{C++} library \textsc{Xsimd}.

\section{Demonstration}

In addition to the software we have developed for the purposes of this paper (Section \ref{sec:software}), we have used the \textsc{R} programming language \citep{rlang} and the \textsc{R} graphics package \textsc{ggplot2} \citep{ggplot} to produce the figures and results in the following.  The 95\% credible intervals we present are highest posterior density intervals that we obtain using the \textsc{R} package \textsc{coda} \citep{coda}.

\subsection{Parallelization}\label{sec:ParResults}

\begin{figure}[t]
	\centering
	\includegraphics[width=0.8\textwidth]{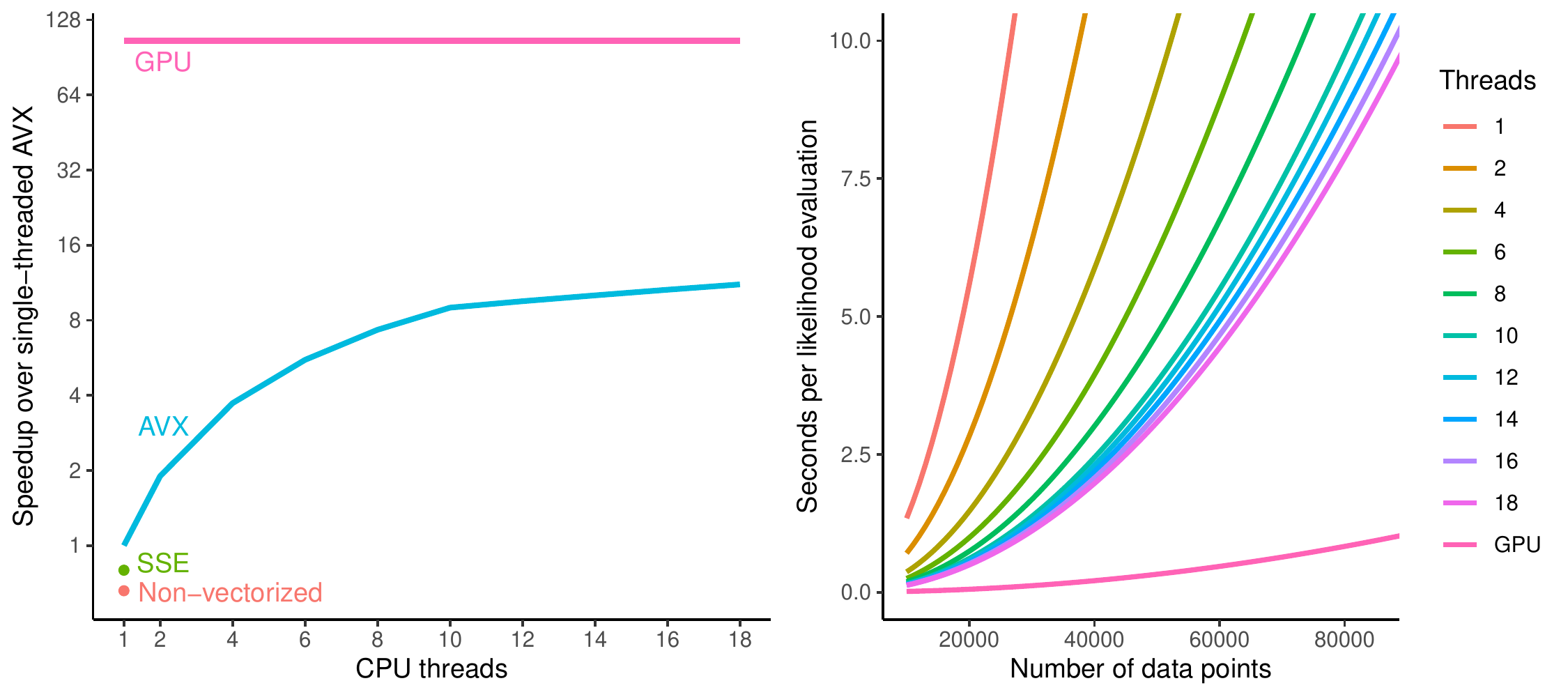}
	\caption{Spatiotemporal Hawkes process likelihood evaluations. [Left] Speedup of graphics processing unit (GPU) and multi-core advanced vector extensions (AVX) computations relative to single-core AVX computing, all using 75,000 randomly generated observations.  Single-core implementations without single instruction, multiple data (SIMD) and with streaming SIMD extensions (SSE) occupy the bottom left corner. [Right] Seconds to compute for both GPU and multi-core AVX processing as a function of data quantity.}
	\label{fig:performFig}
\end{figure}

For CPU results, we use a Linux machine with a 10-core Intel Xeon W-2155 processor (3.3 GHz).  Each core supports 2 independent threads or logical cores, so the machine reaches a peak performance of 264 gigaflops with double-precision floating point enhanced with AVX vectorization (double that for fused operations such as fused multiply-add). The processor comes with 32 GB DDR4 memory (2667 MHz), 640 KB L1 cache, 10 MB L2 cache and 13 MB L3 cache.  For the GPU results, we use an NVIDIA Titan V with 5120 CUDA cores (1.2 GHz), achieving 3.1 teraflops peak double-precision floating point performance (again, double this for fused operations).  The Titan V comes with 12 GB HBM2 memory, and its 5120 CUDA cores divide into 80 separate \emph{streaming multiprocessors} (SM), each consisting of 64 CUDA cores and its own 96 KB L1 cache.  Together, all 80 SMs share a single 4.5 MB L2 cache.

Figure \ref{fig:performFig} shows GPU, single-core, multi-core and vectorized processing performances for spatiotemporal Hawkes process likelihood evaluations. On the left, we randomly generate $N=75,000$ data points and observe relative speedups over single-threaded AVX processing (77.19 $s$). The GPU implementation (0.73 $s$) is 105$\times$ faster and the 18 thread AVX implementation (6.93 $s$) is 10.4$\times$ faster. The roughly 10-fold speedup of the GPU implementation over the 18 thread AVX implementation accords with the former's 3.1 teraflop peak performance relative to the latter's 0.3 teraflop peak performance.  On the other hand, the single-threaded AVX implementation is $1.26\times$ and $1.52\times$ faster than the SSE (96.94 $s$) and non-vectorized (117.16 $s$) implementations, respectively.  Finally, the GPU implementation is 160$\times$ the speed of the single-threaded non-vectorized implementation.  On the right, we observe the number of seconds required to perform a single likelihood evaluation for our different implementations as a function of the number of observations, which we let scale from 10,000 data points to 90,000 data points.  We compare GPU performance to single- and multi-threaded AVX processing.  As expected, all implementations appear to take on a quadratic curve, although one might imagine that the GPU performance has a significantly smaller leading constant.

\subsection{Gunshots in Washington D.C.}

\begin{figure}[t]
	\centering
	\includegraphics[width=0.8\textwidth]{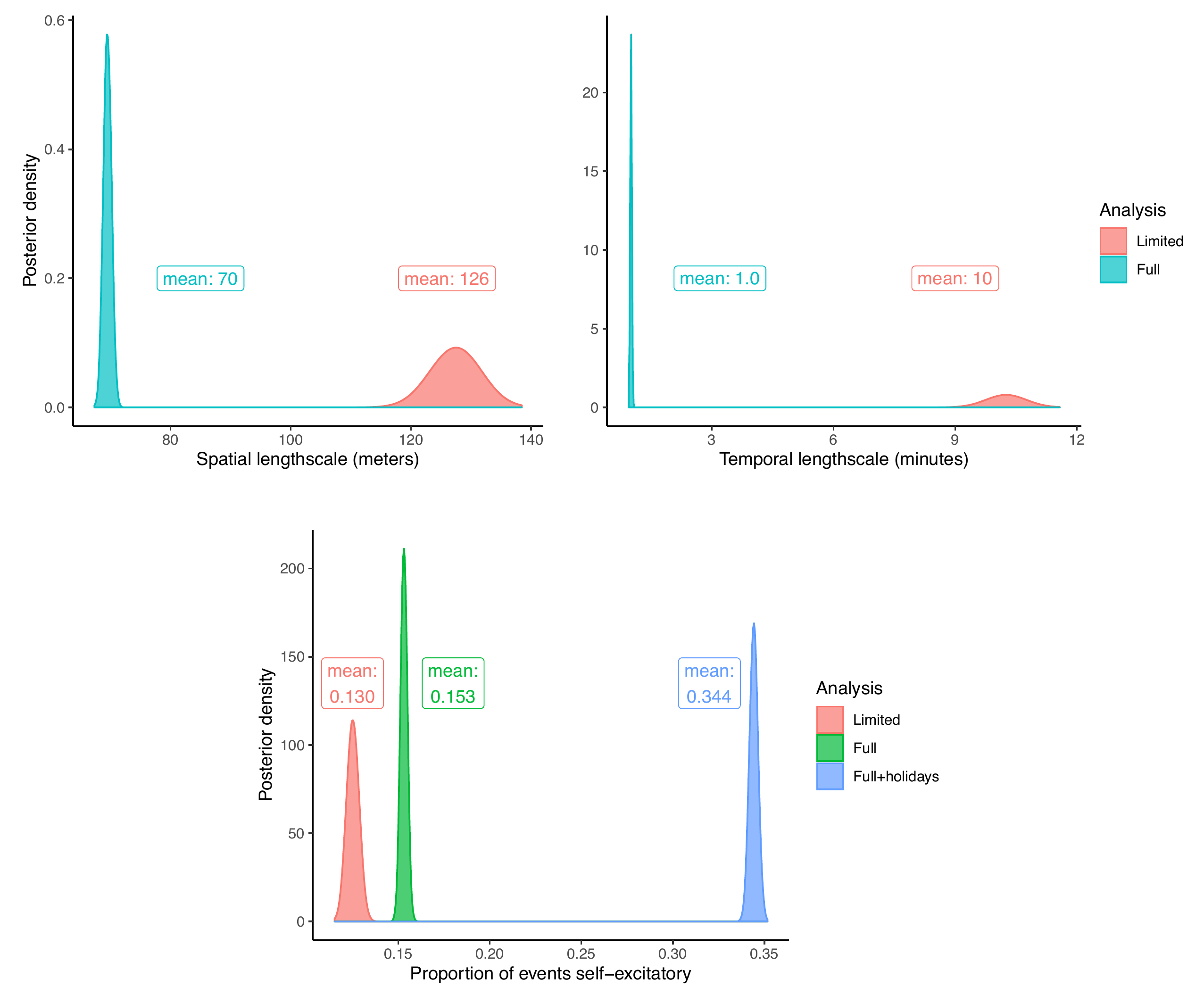}
	\caption{Posterior distributions of model parameters conditioned on different datasets: `limited' indicates the 2010 to 2012 analysis of \cite{loeffler2018gun} (9,000+ observations); `full' indicates the 2006 to 2019 analysis without New Years and July 4 (55,000+ observations); `full+holidays' indicates a 2006 to 2019 analysis including New Years and July 4 (85,000+ observations).  Larger lengthscales for the limited analysis likely result from thinning of events within the same minute and 100 meter range.  Both full and limited proportion of events self-excitatory $(\theta)$ are within the previously estimated range of 10 to 18\%, whereas that of full+holidays is nowhere near previously estimated ranges.}
	\label{fig:densities}
\end{figure}

\begin{figure}[t]
	\centering
	\includegraphics[width=0.9\textwidth]{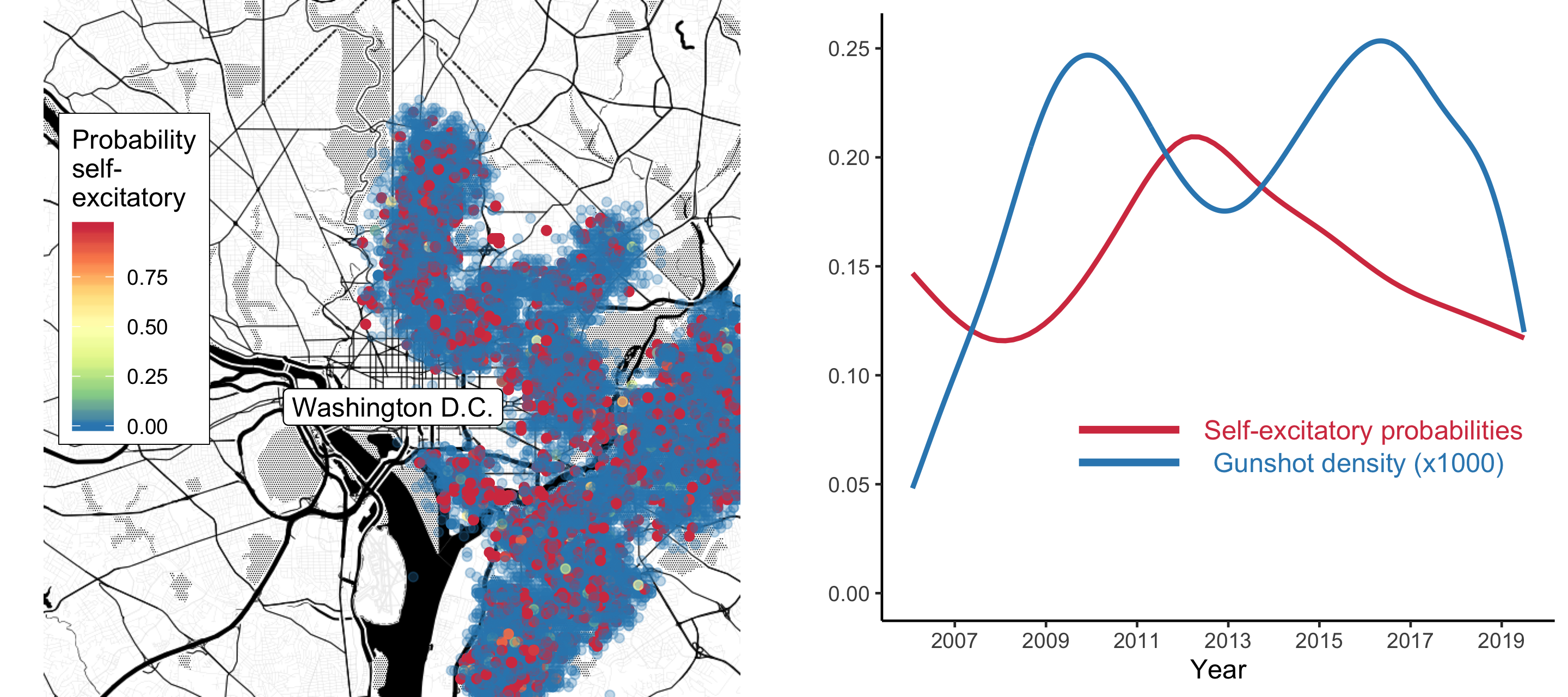}
	\caption{Posterior means for self-excitatory probabilities $\pi_n$ (Equation \eqref{eq:prob_se}) in relation to spatial and temporal allocations. [Left] Red indicates a high posterior probability of a gunshot being self-excitatory in nature; blue indicates a low posterior probability. Few yellow points suggests concentration towards values 0 and 1. [Right] We compare smoothing of posterior means for self-excitatory probabilities as a function of time with empirical gunshot trends.  A peak in the former around 2013 appears to correspond to a nadir for the latter. }
	\label{fig:dcmap}
\end{figure}

\begin{figure}[t]
	\centering
	\includegraphics[width=\textwidth]{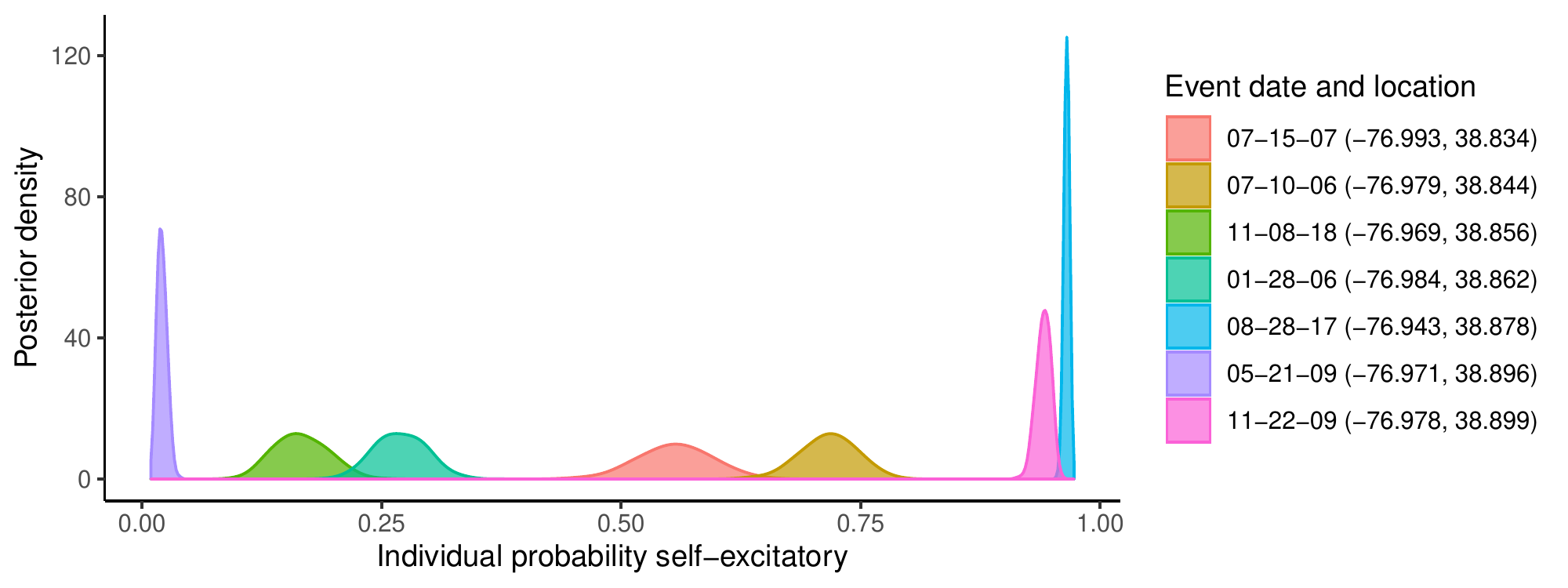}
	\caption{Posterior distributions for 7 individual probabilities $\pi_n$ that each gunshot event results from self-excitation. Such distributions may be useful for ascertaining whether specific instances of gun violence are retaliatory in nature.  As expected, probabilties close to 0 and 1 vary less. The majority of $\pi_n$ (not visualized here) resemble point masses extremely close to 0 or 1.}
	\label{fig:individual_probs}
\end{figure}

We apply our inference framework to AGLS data generated in Washington D.C. between the years 2006 and 2019 to ascertain the nature of gun violence as a collective phenomenon.  Specifically, we wish to determine the extent to which gunfire in D.C.~is contagious or diffusionary in nature.  We build on, and compare our results to, the analysis of \cite{loeffler2018gun}, which uses a similar model to that specified in Section \ref{sec:model}.  That analysis obtained results from 9,000+ data points collected in the years 2010, 2011 and 2012, and the data we use differs from that data two ways.  First, we combine datasets located at \url{justicetechlab.org/shotspotter-data} \citep{carr2016geography,carr2018keep} and \url{https://opendata.dc.gov/datasets} to obtain data from 2006 to 2013 and from 2014 to 2019, respectively.  Second, these data include the exact second of each event and so have greater temporal precision than that of the previous analysis, which considered data points within the same minute and 100 meter radius to be duplicates.  In this way, the current dataset is larger because of both greater temporal breadth and greater temporal precision.  Like the previous analysis, we consider two datasets, one with all days of the year and one with New Year's Eve, July 4 and surrounding days removed on account of false positives from fireworks and celebratory gunfire.
The former (`full+holidays') consists of 85,000+ observations, the latter (`full') 55,000+ observations.

We use Algorithm \ref{alg:mh} to generate 4 Markov chains of 10,000 states each and discard the first 1,000 states of each chain.  Using our GPU and Algorithm \ref{alg:lik} to calculate the likelihood within the accept-reject step, total compute time lasts about 4 hours for the full analysis and 10 hours for the full+holidays analysis.  Effective sample sizes are greater than 1,700 for all parameters.  The top row of Figure \ref{fig:densities} compares posterior inference for lengthscale parameters $1/\omega$, the temporal lengthscale, and $h$, the spatial lengthscale, between the `limited' analysis of \cite{loeffler2018gun} and our full analysis.  We obtain posterior means of 69.5 meters (95\% CI: 68.5, 70.8) and 1.0 minute (95\% CI: 0.98, 1.04) for the two lengthscales, compared to 126 meters (95\% CI: 121, 134) and 10 minutes (95\% CI: 9.5, 11) for the limited analysis. Both of these results may be expected because the limited analysis removed events within the same minute and 100 meters to obtain a thinned dataset about 95\% of the original size.  As a result, we estimate retaliatory gunfire to occur much sooner after, and closer to, a previous gunshot. To verify this trend, we perform a sensitivity test and remove 8\% of the full dataset by considering events within a minute and 100 meters from each other to be duplicates. This sensitivity test results in posterior means of 262.4 meters (95\% CI: 253.3, 270.7) and 46.2 minutes (95\% CI: 43.6, 48.7) for the two lengthscales. Further sensitivity tests based on 15\% and 20\% thinned datasets revealed even larger lengthscales.  Returning to the full analysis, the posterior variances arising from the full analysis are significantly smaller.  This makes sense for two reasons: first, the data conditioned upon are over 5$\times$ larger; second, we are considering positive random variables, the variance of which scales with the mean.

In the second row of Figure \ref{fig:densities}, we compare posterior densities for parameter $\theta$, which represents the relative weight of the background intensity or the general proportion of events that are self-excitatory in nature.  Here, the posterior mean of $\theta$ conditioned on the full dataset is 0.153 (95\% CI: 0.150, 0.156) and larger than the 0.13 (95\% CI: 0.12, 0.13) of the limited analysis.  Again, we attribute this to the lack of data thinning in the full dataset and the resulting greater temporal proximity of gunshots, but we note that both posterior densities nest well within the estimated range of 10 to 19\% for retaliatory homicides of \cite{juvenileHomicide}.  On the other hand, the posterior mean of the full+holidays analysis is artificially inflated to 0.344 (95\% CI: 0.340, 0.348) by cascades of fireworks and celebratory gunfire encompassing over one-third of that dataset.

The second half of our analysis considers posterior distributions for the probabilities $\pi_n$ of each individual gunshot event arising from self-excitation (i.e., being retaliatory) as opposed to the background process.  We use our GPU to apply Algorithm \ref{alg:pi} to---for storage reasons---a thinned sample of 1,000 $\Theta^{(s)}$s to produce 1,000 vectors $\pi^{(s)}$ each of length 55,000.  In Figure \ref{fig:dcmap}, we visualize the distribution of the posterior means in space and time.  On the left, red self-excitatory events distribute fairly evenly among blue background events in Washington D.C., while yellow neutral events barely exist. As a sanity check, the proportion of self-excitatory events seems to roughly coincide with the estimated 0.15 posterior mean of $\theta$.  On the right, we smooth posterior self-excitatory probabilities for each event through time, from 2006 to 2009, and compare to the overall gunshot density.  In general,
the trend in self-excitatory events hits a peak in 2013 of about 20\%.  This peak coincides with a small dip in the total gunshots for the year 2013, indicating fewer and more closely connected gunfire clusters.  Censoring issues make it difficult to interpret relations in these trends near 2006 and 2019.  Finally, Figure \ref{fig:individual_probs} presents posterior distributions of probabilities $\pi_n$ that 7 individual events are self-excitatory in nature.  As may be inferred from Figure \ref{fig:dcmap}'s few yellow points, most events cluster close to 0 or 1, resembling a point mass.  But many events do provide significant uncertainty, and, as expected, those with posterior mean closer to 0.5 have much greater variability.  We believe that figures like Figure \ref{fig:individual_probs} may be useful for crime investigations in determining the retaliatory nature of specific acts of gun violence and quantifying uncertainty in this regard.

\textsc{R} code, data and posterior samples related to the above analysis are available at \url{https://github.com/andrewjholbrook/shot_spotter}.  We point out that spatial and temporal censoring bias our results, and we consider corrections for, and modeling of, such bias in a big data context to be a fascinating next step in this line of research.

\section{Discussion}

Self-excitatory stochastic process models are useful for modeling complex diffusionary and cascading phenomena in multiple scientific disciplines and industrial sectors, but the computational complexity of statistical inference for these models has barred them from applications involving big data.  In this paper, we have developed a high-performance statistical computing framework for Hawkes process models that leverages contemporary computational hardware and scales Bayesian inference to more thant 85,000 observations. To accomplish this, we have created software for both vectorized multi-core CPU and many-core GPU architecture implementations and made this open-source software freely available online.  As a demonstration of the usefulness of this approach, we have applied a spatiotemporal Hawkes process model to the analysis of emerging acoustic gunshot locator systems data recorded in the neighborhoods of Washington D.C.~between the years of 2006 and 2019.  In this context, Bayesian inference facilitated by our framework provided point estimation and uncertainty quantification of the nature of gun violence as a contagion in American communities.  To this end, we have created an additional massively parallel post-processing pipeline to compute probabilities that individual events result from self-excitation based on posterior samples arising from MCMC.  These posterior probabilities have proven useful for creating spatial and temporal visualizations that relate self-excitatory gun violence to the Washington D.C. landscape and for quantifying our uncertainty whether individual events are retaliatory in origin.  We hope this analysis brings attention to big, complex and emeging AGLS data, the analysis of which might improve scientific understanding of the great American gun violence epidemic.

In the context of this poorly understood epidemic in which many complex models might be posited, fast inference is all the more necessary to facilitate quick candidate model comparison.  For example, it is highly doubtful that all self-excitatory action is purely retaliatory in nature: shooting events may consist of multiple shots by the same individual or group. On the other hand, retaliatory shootings may plausibly occur days, weeks or even months after a precipitating event. Thus, it seems that a mixture model employing multiple triggering functions would be appropriate to combine a very short time frame with a slightly longer one or with, perhaps, a much larger time variation (days to months).
The reality of multi-shot shooting events, very short-term gunfights and longer term retaliation occurring minutes, hours, days or even months later suggests that additional models to capture these different processes operating over multiple spatial and temporal scales will be needed. This only reinforces the need for fast computation, which will support selection between more complex models as well as comparisons to the simpler ones already in the literature.

We will also extend our high-performance computing framework to other generalizations of the Hawkes process such as marked Hawkes processes and mutually-exciting point processes.  The former have been effective for modeling Earthquakes (here, the mark is the tremor's score on the Richter scale), the latter for modeling dependencies between neurons.  For these efforts to succeed and enjoy maximal impact, we must scale Bayesian inference for such point process models to millions of observations, and we believe that computational tools that accomplish fine-grained parallelization (e.g.~tensor processing units and bigger, faster GPUs) will accomplish more than multi-processor approaches that fail to overcome inherent latency and communication bottlenecks.  Nonetheless, we are also interested in developing inference frameworks that share computational resources between both CPU and GPU simultaneously.  For scalable Bayesian inference, all computing tools and computational hardware must be on the table. After all, Washington D.C. is only one city of at least 40 for which AGLS data have come available in the last decade: American gunfire data are big data, indeed.

\section*{Acknowledgments}

The research leading to these results has received funding through National Institutes of Health grant U19 AI135995 and National Science Foundation grant DMS1264153.
We gratefully acknowledge support from NVIDIA Corporation with the donation of parallel computing resources used for this research.

\bibliographystyle{sysbio}
\bibliography{refs}

\end{document}